\documentclass[preprint,amsmath,amssymb,aps,showkeys,showpacs]{revtex4}
\usepackage[english]{babel}
\usepackage{graphicx}
\usepackage{graphics}
\usepackage{amsmath}
\usepackage{dcolumn}
\usepackage{amssymb}
\usepackage{bm}
\usepackage[latin1]{inputenc}

\begin{document}
\title{On a relativistic scalar particle subject to a Coulomb-type potential given by Lorentz symmetry breaking effects}
\author{K. Bakke}
\email{kbakke@fisica.ufpb.br}
\affiliation{Departamento de F\'isica, Universidade Federal da Para\'iba, Caixa Postal 5008, 58051-970, Jo\~ao Pessoa, PB, Brazil.}

\author{H. Belich} 
\email{belichjr@gmail.com}
\affiliation{Departamento de F\'isica e Qu\'imica, Universidade Federal do Esp\'irito Santo, Av. Fernando Ferrari, 514, Goiabeiras, 29060-900, Vit\'oria, ES, Brazil.}

\begin{abstract}
The behaviour of a relativistic scalar particle in a possible scenario that arises from the violation of the Lorentz symmetry is investigated. The background of the Lorentz symmetry violation is defined by a tensor field that governs the Lorentz symmetry violation out of the Standard Model Extension. Thereby, we show that a Coulomb-type potential can be induced by Lorentz symmetry breaking effects and bound states solutions to the Klein-Gordon equation can be obtained. Further, we discuss the effects of this Coulomb-type potential on the confinement of the relativistic scalar particle to a linear confining potential by showing that bound states solutions to the Klein-Gordon equation can also be achieved, and obtain a quantum effect characterized by the dependence of a parameter of the linear confining potential on the quantum numbers $\left\{n,l\right\}$ of the system.
\end{abstract}


\maketitle

\section{Introduction}

The Higgs particle, a cornerstone of the standard model finally has been detected. This fact closes a cycle of research that has initiated in the beginning of twenty century and it closes a scenario of forecasts that have a great experimental success. Despite this relevant fact, there is a lack of a more fundamental theory that covers the description neutrino without mass and can also explain the microscopic origin of the boson which generates the mass of all the particles of the universe. Another intriguing point of discussion in the area of particle physics is the origin of electron electric dipole moment which has not been explained by Standard Model of particle physics yet. At present days, it is well-known that just experimental upper bounds have been established \cite{revmod}. Based on the Standard Model, an upper limit for electron electric dipole moment has been established as $d_{e}\leq 10^{-38}\,\mathrm{e}\cdot\mathrm{cm}$ \cite{revmod}. On the other hand, experiments measured an upper limit given by $d_{e}\leq 10^{-29}\,\mathrm{e}\cdot\mathrm{cm}$ by using a polar molecule thorium monoxide (ThO) \cite{science}. This experimental result has shown us a necessity of investigating the physics beyond the Standard Model because the term associated with the electric dipole moment violates the CP symmetry.

A possible way of dealing with a scenario beyond the Standard Model is the extension of the mechanism for the spontaneous symmetry breaking through vector or tensor fields, which implies that the Lorentz symmetry is violated. This scenario has been established after the seminal work made by Kosteleck\'y and Samuel \cite{extra3} in the string theory. It is shown that the Lorentz symmetry is violated through a spontaneous symmetry breaking mechanism triggered by the appearance of nonvanishing vacuum expectation values of nontrivial Lorentz tensors. Such models that deal with the physics beyond the Standard Model are considered as effective theories, whose analysis of the phenomenological aspect at low energies can provide information and impose restrictions on the fundamental theory in which they stem from. A general framework for testing the low-energy manifestations of the CPT symmetry and the Lorentz symmetry breaking is known as the Standard Model Extension (SME) \cite{colladay-kost}. In this framework, the effective Lagrangian operator corresponds to the usual Lagrangian operator of the Standard Model to which is added to the Standard Model operators a Lorentz violation tensor background. The effective Lagrangian is written as being an invariant under the Lorentz transformation of coordinates in order to guarantee that the observer independence of physics. However, the physically relevant transformations are those that affect only the dynamical fields of the theory. These changes are called as particle transformations, whereas the coordinate transformations (including the tensor background) are called as the observer transformations. In Refs. \cite{coll-kost,baeta,bras}, one can find a deep analysis of these concepts. Concerning the experimental searches for the CPT/Lorentz-violation signals, the generality of the SME has provided the basis for many investigations. In the flat spacetime limit, empirical studies include muons \cite{muon}, mesons \cite{meson,meson2}, baryons \cite{barion,barion2}, photons \cite{photon,kost2}, electrons \cite{electron}, neutrinos \cite{neutrino} and the Higgs sector \cite{higgs}. The gravity sector has also been explored in Refs. \cite{gravity,gravity2,bb15}. In Ref. \cite{data}, one can find the current limits on the coefficients of the Lorentz symmetry violation. In recent years, Lorentz symmetry breaking effects have been investigated in the hydrogen atom \cite{manoel}, in Weyl semi-metals \cite{weyl}, on the Rashba coupling \cite{rash,bb3}, in the quantum Hall effect \cite{lin2}, in tensor backgrounds \cite{louzada,manoel2} and geometric quantum phases \cite{belich,belich1,bb2,bb4,lbb}.

In this paper, we study a relativistic scalar particle in a possible scenario of anisotropy generated by a Lorentz symmetry breaking term defined by a tensor $\left( K_{F}\right)_{\mu\nu\alpha\beta}$ that corresponds to a tensor that governs the Lorentz symmetry violation out of the Standard Model Extension. We investigate the effects of a Coulomb-type potential induced by a Lorentz symmetry violation background on a relativistic scalar particle by showing that bound states solutions to the Klein-Gordon equation can be obtained in a particular scenario of the Lorentz symmetry violation. Further, we investigate the effects of a Coulomb-type potential induced by a Lorentz symmetry violation background on the confinement of the relativistic scalar particle to a linear confining potential and obtain a quantum effect characterized by the dependence of parameter $\eta$ of the linear confining potential on the quantum numbers $\left\{n,l\right\}$ of the system, whose meaning is that not all values of the parameter $\eta$ are allowed in order to obtain the bound states solutions to the Klein-Gordon equation.

The structure of this paper is as follows: in section II, we introduce a background of the Lorentz symmetry violation defined by a tensor $\left( K_{F}\right)_{\mu\nu\alpha\beta}$ that governs the Lorentz symmetry violation out of the Standard Model Extension; thus, we establish a possible scenario of the Lorentz symmetry violation that gives rise to a Coulomb-type potential and solve the Klein-Gordon equation; in section III, we confine the relativistic scalar particle to a linear scalar potential, and discuss the effects of the Coulomb-type potential that stems from Lorentz symmetry breaking effects on this confinement; in section IV, we present our conclusions.

\section{Background of the Lorentz symmetry violation}

Spontaneous symmetry violation has become an important ingredient in the search for fundamental theories that can incorporate a unified view of physics. Examples of the spontaneous symmetry breaking are phase transitions by a scalar field and the Higgs mechanism \cite{ryder}, which, in turn, have clarified the origin of the inertial mass of the particles. An interesting perspective of investigation is to analyze the effects of anisotropy in the phase transition in which the spontaneous breaking of symmetry has already taken place, and see how these effects can affect the physical properties of a particle immersed in this background. In this section, we deal with a relativistic scalar particle subject to an anisotropic environment generated by a violating term of the Lorentz symmetry.

The gauge sector of the Standard Model Extension presents two violating terms that modifies the transport properties of the spacetime: the CPT-odd sector \cite{colladay-kost,coll-kost} and the CPT-even sector \cite{baeta,jackiw}. An interesting property is the vacuum birefringence yielded by the SME \cite{coll-kost,baeta}. Our goal is to investigate how a scalar field can be influenced by the anisotropy of the spacetime generated by background fields that arises from a soft spontaneous breaking of the Lorentz symmetry. As it is expected, terms that violate the Lorentz symmetry start with tiny values and go for energy scales beyond the Standard Model. For this purpose, we relax the renormalization property of our model. Therefore, despite being inspired by the CPT-even sector of the SME, our proposal is not inserted in the scenario of SME. This is promising because it can open new discussions about new theories that covers energy scales beyond the SME. In order to investigate this possibility, let us study the behaviour of a relativistic scalar particle in the presence of effects that favor a determined direction. Thereby, inspired by Refs. \cite{colladay-kost,coll-kost,kost2,louzada}, let us write the Klein-Gordon equation as:
\begin{eqnarray}
p^{\mu}p_{\mu}\phi-\frac{g}{4}\,\left(K_{F}\right)_{\mu\nu\alpha\beta}\,F^{\mu\nu}\left(x\right)\,F^{\alpha\beta}\left(x\right)\phi=m^{2}\,\phi
\label{1.1}
\end{eqnarray}
where $g$ is a constant, $\left(K_{F}\right)_{\mu\nu\alpha\beta}$ corresponds to a tensor that governs the Lorentz symmetry violation out of the Standard Model Extension \cite{colladay-kost,coll-kost,baeta,kost2}. In particular, the tensor $\left(K_{F}\right)_{\mu\nu\alpha\beta}$ can be written in terms of four $3\times3$ matrices defined as follows:
\begin{eqnarray}
\left(\kappa_{DE}\right)_{ij}&=&-2\left(K_{F}\right)_{0j0k};\nonumber\\
\left(\kappa_{HB}\right)_{jk}&=&\frac{1}{2}\epsilon_{jpq}\,\epsilon_{klm}\left(K_{F}\right)^{pqlm}\\
\left(\kappa_{DB}\right)_{jk}&=&-\left(\kappa_{HE}\right)_{kj}=\epsilon_{kpq}\left(K_{F}\right)^{0jpq}.\nonumber
\label{1.2}
\end{eqnarray}

The matrices $\left(\kappa_{DE}\right)_{ij}$ and $\left(\kappa_{HB}\right)_{ij}$ are symmetric and represent the parity-even sector of the tensor $\left(K_{F}\right)_{\mu\nu\alpha\beta}$. On the other hand, the matrices $\left(\kappa_{DB}\right)_{ij}$ and $\left(\kappa_{HE}\right)_{ij}$ has no symmetry, and represent the parity-odd sector of the tensor $\left(K_{F}\right)_{\mu\nu\alpha\beta}$. Hence, by writing the tensor $\left(K_{F}\right)_{\mu\nu\alpha\beta}$ in terms of the matrices given in Eq. (\ref{1.2}), we can rewrite the Klein-Gordon equation in the form \cite{louzada}:
\begin{eqnarray}
m^{2}\,\phi=p^{\mu}p_{\mu}\phi+\frac{g}{2}\left(\kappa_{DE}\right)_{i\,j}\,E^{i}\,E^{j}\,\phi-\frac{g}{2}\,\left(\kappa_{HB}\right)_{j\,k}\,B^{i}\,B^{j}\,\phi+g\left(\kappa_{DB}\right)_{j\,k}\,E^{i}\,B^{j}\,\phi.
\label{1.3}
\end{eqnarray}

In this section, we establish a possible scenario of the Lorentz symmetry violation that gives rise to a Coulomb-type potential; thus, we show that bound states solutions of the Klein-Gordon equation (\ref{1.3}) can be obtained. Thereby, let us consider a field configuration defined as
\begin{eqnarray}
\vec{E}=\frac{\lambda}{\rho}\,\hat{\rho};\,\,\,\,\,\vec{B}=B_{0}\,\hat{z},
\label{1.4}
\end{eqnarray}
where $B_{0}>0$ and $\lambda$ are constants, $\hat{\rho}$ and $\hat{z}$ are unit vectors in the radial and $z$-direction, respectively. Let us also consider only one non-null component of the tensor $\left(K_{F}\right)_{\mu\nu\alpha\beta}$ given by 
\begin{eqnarray}
 \left(\kappa_{DB}\right)_{13}=\kappa=\mathrm{const}.
\label{1.4a}
\end{eqnarray}

Hence, from now on, we work with a Lorentz symmetry breaking scenario given by field configuration given in Eq. (\ref{1.4}) and the tensor, whose components are defined in Eq. (\ref{1.4a}). In this way, by writing the Minkowski spacetime in the form:
\begin{eqnarray}
ds^{2}=-dt^{2}+d\rho^{2}+\rho^{2}\,d\varphi^{2}+dz^{2},
\label{1.5}
\end{eqnarray}
we can write the Klein-Gordon equation (\ref{1.3}) as
\begin{eqnarray}
m^{2}\,\phi=-\frac{\partial^{2}\phi}{\partial t^{2}}+\frac{\partial^{2}\phi}{\partial\rho^{2}}+\frac{1}{\rho}\frac{\partial\phi}{\partial\rho}+\frac{1}{\rho^{2}}\frac{\partial^{2}\phi}{\partial\varphi^{2}}+\frac{\partial^{2}\phi}{\partial z^{2}}+\frac{g\,\kappa\,\lambda\,B_{0}}{\rho}\,\phi
\label{1.5}
\end{eqnarray}

In what follows, let us consider a particular solution to Eq. (\ref{1.5}) given by the eigenfunctions of the operators $\hat{L}_{z}=-i\partial_{\varphi}$ and $\hat{p}_{z}=-i\partial_{z}$. Since these operators commute with the Hamiltonian operator given in the right-hand-side of Eq. (\ref{1.5}), then, we can write the particular solution to Eq. (\ref{1.5}) in terms of the eigenvalues of the $z$-component of the angular momentum $\hat{L}_{z}=-i\partial_{\varphi}$ and the operator $\hat{p}_{z}=-i\partial_{z}$ as follows:
\begin{eqnarray}
\phi=e^{-i\mathcal{E}t}\,e^{il\varphi}\,e^{ikz}\,f\left(\rho\right),
\label{1.6}
\end{eqnarray}
 where $l=0,\pm1,\pm2,\ldots$, $k$ is a constant and $f\left(\rho\right)$ is a function of the radial coordinate. Henceforth, we consider $k=0$. Then, substituting Eq. (\ref{1.6}) into Eq. (\ref{1.5}), we obtain
\begin{eqnarray}
\frac{d^{2}f}{d\rho^{2}}+\frac{1}{\rho}\frac{df}{d\rho}-\frac{l^{2}}{\rho^{2}}\,f+\frac{\delta}{\rho}\,f-\tau^{2}\,f=0,
\label{1.7}
\end{eqnarray}
where we have defined the following parameters:
\begin{eqnarray}
\tau^{2}&=&m^{2}-\mathcal{E}^{2};\nonumber\\
[-2mm]\label{1.8}\\[-2mm]
\delta&=&g\kappa\,\lambda\,B_{0}.\nonumber
\end{eqnarray}

Note that the fourth term of Eq. (\ref{1.7}) plays the role of a Coulomb potential. It behaves like an attractive potential if $\delta>0$. This is possible if $\lambda>0$. Let us perform a change of variables given by $\xi=2\tau\rho$. Then, we have
\begin{eqnarray}
\frac{d^{2}f}{d\xi^{2}}+\frac{1}{\xi}\frac{df}{d\xi}-\frac{l^{2}}{\xi^{2}}\,f+\frac{\delta}{2\tau\xi}\,f-\frac{1}{4}\,f=0.
\label{1.10}
\end{eqnarray}
By analysing the asymptotic behaviour of Eq. (\ref{1.10}), we have that a solution to Eq. (\ref{1.10}) can be written in terms of an unknown function $F\left(\xi\right)$ as: 
\begin{eqnarray}
f\left(\xi\right)=e^{-\frac{\xi}{2}}\,\xi^{\left|l\right|}\,F\left(\xi\right).
\label{1.11}
\end{eqnarray}
Thereby, substituting Eq. (\ref{1.11}) into Eq. (\ref{1.10}) we obtain
\begin{eqnarray}
\xi\,\frac{d^{2}F}{d\xi^{2}}+\left[2\left|l\right|+1-\xi\right]\frac{dF}{d\xi}+\left[\frac{\delta}{2\tau}-\left|l\right|-\frac{1}{2}\right]F=0,
\label{1.12}
\end{eqnarray}
which is called as the confluent hypergeometric equation \cite{abra,arf}; thus, the function $F\left(\xi\right)$ is the confluent hypergeometric function, that is, $F\left(\xi\right)=\,_{1}F_{1}\left(\left|l\right|+\frac{1}{2}-\frac{\delta}{2\tau},\,2\left|l\right|+1,\,\xi\right)$. It is well-known that the confluent hypergeometric series becomes a polynomial when $\left|l\right|+\frac{1}{2}-\frac{\delta}{2\tau}=-n$ \cite{arf,abra}, where $n=0,1,2,\ldots$. In this way, we have
\begin{eqnarray}
\mathcal{E}_{n,\,l}=\pm\,\sqrt{m^{2}-\frac{\delta^{2}}{4\left[n+\left|l\right|+\frac{1}{2}\right]^{2}}}\,.
\label{1.13}
\end{eqnarray}

Hence, Eq. (\ref{1.13}) are the allowed energy of this relativistic system yielded by the effects of a Coulomb-type potential induced by a Lorentz symmetry violation background out of the Standard Model Extension on a relativistic scalar particle. In the present case, we have established a particular background of the violation of the Lorentz symmetry defined by tensor background that governs the Lorentz symmetry violation possessing a non-null component given by $\left(\kappa_{DB}\right)_{13}=\kappa=\mathrm{const}$, a radial electric field produced by a linear distribution of electric charges on the $z$-axis and a uniform magnetic along the $z$-direction. Note that a change of the background of the Lorentz symmetry violation does not allow us to obtain the bound states solutions given above. For instance, by considering the non-null component of the tensor  as $\left(\kappa_{DB}\right)_{11}=\mathrm{const}$, then, the spectrum of energy given in Eq. (\ref{1.13}) cannot be obtained because the Coulomb-type potential that appears in Eq. (\ref{1.7}) would not exist.

\section{confinement to a linear scalar potential}

In this section, we discuss the effects of the Lorentz symmetry violation when a relativistic scalar particle is subject to a scalar potential. The most known procedure in introducing a scalar potential into the Klein-Gordon equation follows the same procedure in introducing the electromagnetic 4-vector potential \cite{greiner}. This introduction occurs by modifying the momentum operator $p_{\mu}=i\partial_{\mu}$ in the form: $p_{\mu}\rightarrow p_{\mu}-q\,A_{\mu}\left(x\right)$. In recent years, the confinement of a relativistic scalar particle to a Coulomb potential has been discussed by several authors \cite{kg,kg2,kg3,kg4,greiner}. On the other hand, it is shown in  Ref. \cite{scalar} that the scalar potential can also been introduced into the Klein-Gordon equation by modifying in the mass term as follows: $m\rightarrow m+S\left(\vec{r},\,t\right)$, where $S\left(\vec{r},\,t\right)$ is the scalar potential. Examples of this second procedure are the study of he behaviour of a Dirac particle in the presence of static scalar potential and a Coulomb potential \cite{scalar2}, and a relativistic scalar particle in the cosmic string spacetime \cite{eug}. 

In this work, we introduce the scalar potential by modifying the mass term of the Klein-Gordon equation and analyse the effects of the Coulomb-type potential that stems from the effects of the Lorentz symmetry violation on the confinement of a relativistic scalar potential to this scalar potential. Thereby, let us consider a scalar confining potential given by
\begin{eqnarray}
S\left(\rho\right)=\eta\,\rho,
\label{2.1}
\end{eqnarray}
where $\eta$ is a constant. The linear scalar potential given in Eq. (\ref{2.1}) has been proposed to describe the confinement of quarks \cite{linear} due to experimental data show a behaviour of the confinement to proportional to the distance between the quarks \cite{linear4}. It has also been explored in studies of the quark-antiquark interaction as a problem of a relativistic spinless particle which possesses a position-dependent mass, where the mass term acquires a contribution given by a interaction potential that consists of a linear and a harmonic confining potential plus a Coulomb potential term \cite{bah}. Besides, the linear scalar potential given in Eq. (\ref{2.1}) has attracted a great interest in atomic and molecular physics as pointed out in Ref. \cite{linear3} and in several discussions of relativistic quantum mechanics \cite{linear2,eug,scalar,scalar2,vercin,mhv}. Therefore, the Klein-Gordon equation (\ref{1.5}) becomes
\begin{eqnarray}
\left[m+S\left(\rho\right)\right]^{2}\phi=-\frac{\partial^{2}\phi}{\partial t^{2}}+\frac{\partial^{2}\phi}{\partial\rho^{2}}+\frac{1}{\rho}\frac{\partial\phi}{\partial\rho}+\frac{1}{\rho^{2}}\frac{\partial^{2}\phi}{\partial\varphi^{2}}+\frac{\partial^{2}\phi}{\partial z^{2}}+\frac{g\,\kappa\,\lambda\,B_{0}}{\rho}\,\phi
\label{2.2}
\end{eqnarray}
Again, a particular solution to Eq. (\ref{2.2}) is given in Eq. (\ref{1.6}); thus, we can write
\begin{eqnarray}
f''+\frac{1}{\rho}\,f'-\frac{l^{2}}{\rho^{2}}\,f+\frac{\delta}{\rho}\,f-2m\,\eta\,\rho\,f-\eta^{2}\,\rho^{2}\,f+\beta^{2}\,f=0,
\label{2.3}
\end{eqnarray}
where we have defined the parameters $\delta$ and $\beta$ in Eq. (\ref{1.8}). Now, let us consider $\lambda>0$, then, we have $\delta>0$ in Eq. (\ref{2.3}). Let us perform a change of variables given by $\zeta=\sqrt{\eta}\,\rho$. Thereby, the radial equation (\ref{2.3}) becomes
\begin{eqnarray}
f''+\frac{1}{\zeta}\,f'-\frac{l^{2}}{\zeta^{2}}\,f+\frac{\delta}{\sqrt{\eta}\rho}\,f-\frac{2m}{\sqrt{\eta}}\,\zeta\,f-\zeta^{2}\,f+\frac{\beta^{2}}{\eta}\,f=0;
\label{2.4}
\end{eqnarray}
thus, a possible solution to Eq. (\ref{2.4} is given in the form:
\begin{eqnarray}
f\left(\zeta\right)=e^{-\frac{\zeta^{2}}{2}}\,e^{\frac{m}{\sqrt{\eta}}\,\zeta}\,\zeta^{\left|l\right|}\,H\left(\zeta\right),
\label{2.5}
\end{eqnarray}
where $H\left(\zeta\right)$ is an unknown function. By substituting Eq. (\ref{2.5}) into Eq. (\ref{2.4}), we obtain the following second order differential equation
\begin{eqnarray}
H''+\left[\frac{2\left|l\right|+1}{\zeta}-\frac{2m}{\sqrt{\eta}}-2\zeta\right]H'-\left[\frac{2m\left|l\right|+m-\delta}{\sqrt{\eta}\,\zeta}\right]H+\left[\frac{\beta^{2}+m^{2}}{\eta}-2-2\left|l\right|\right]H=0,
\label{2.6}
\end{eqnarray}
which is called as the biconfluent Heun equation \cite{heun,eug,bm} and the function $H\left(\zeta\right)$ is the biconfluent Heun function: $H\left(\zeta\right)=H_{B}\left(2\left|l\right|,\,\frac{2m}{\sqrt{\eta}},\,\frac{\beta^{2}+m^{2}}{\eta},-\frac{2\delta}{\sqrt{\eta}},\,\zeta\right)$.

Our focus is on the bound states solutions, therefore, let us use the Frobenius method \cite{arf,eug,bm} from now on. In this way, the solution to Eq. (\ref{2.6}) can be written as a power series expansion around the origin:
\begin{eqnarray}
H\left(\zeta\right)=\sum_{j=0}^{\infty}\,a_{j}\,\zeta^{j}.
\label{2.7}
\end{eqnarray} 

Substituting the series (\ref{2.7}) into (\ref{2.6}), we obtain the recurrence relation:
\begin{eqnarray}
a_{j+2}=\frac{2m\left(j+1\right)+h}{\sqrt{\eta}\left(j+2\right)\,\left(j++2+2\left|l\right|\right)}\,a_{j+1}-\frac{\left(g-2j\right)}{\left(j+2\right)\,\left(j+2+2\left|l\right|\right)}\,a_{j},
\label{2.8}
\end{eqnarray}
where $h=2m\left|l\right|+m-\delta$ and $g=\frac{\beta^{2}+m^{2}}{\eta}-2-2\left|l\right|$. By starting with $a_{0}=1$ and using the relation (\ref{2.8}), we can calculate other coefficients of the power series expansion (\ref{2.7}). For instance,
\begin{eqnarray}
a_{1}&=&\frac{h}{\sqrt{\eta}\left(2\left|l\right|+1\right)};\nonumber\\
[-2mm]\label{2.9}\\[-2mm]
a_{2}&=&\frac{\left(2m+h\right)\,h}{2\eta\left(2+2\left|l\right|\right)\left(2\left|l\right|+1\right)}-\frac{g}{2\left(2+2\left|l\right|\right)}\nonumber.
\end{eqnarray}

In quantum theory, it is required that the wave function of a particle must be normalizable. We assume that the function $f\left(\zeta\right)$ given in Eq. (\ref{2.5}) vanishes as $\zeta\rightarrow0$ and $\zeta\rightarrow\infty$. Hence, the meaning of this condition is that the wave function is finite everywhere, that is, there is no divergence of the wave function as $\zeta\rightarrow0$ and $\zeta\rightarrow\infty$. Bound state solutions correspond to finite solutions, therefore, we can obtain bound state solutions by imposing that the power series expansion (\ref{2.7}) or the biconfluent Heun series becomes a polynomial of degree $n$. Through the expression (\ref{2.8}), we can see that the power series expansion (\ref{2.7}) becomes a polynomial of degree $n$ if we impose the conditions:
\begin{eqnarray}
g=2n\,\,\,\,\,\,\mathrm{and}\,\,\,\,\,\,a_{n+1}=0,
\label{2.10}
\end{eqnarray}
where $n=1,2,3,\ldots$, and $g=\frac{\beta^{2}+m^{2}}{\eta}-2-2\left|l\right|$. As a consequence of the condition $g=2n$, we obtain a discrete spectrum of energy given by:
\begin{eqnarray}
\mathcal{E}_{n,\,l}^{2}=2\,\eta\left[n+\left|l\right|+1\right].
\label{2.11}
\end{eqnarray}

Equation (\ref{2.11}) is the energy levels of a relativistic scalar particle subject to a linear confining potential and under the influence of a Coulomb-type potential yielded by Lorentz symmetry breaking effects, where we have coupled this scalar potential as a modification of the mass term in the Klein-Gordon equation. We can observe that the relativistic energy levels seem to be independent of the parameter $\delta$ associated with the violation of the Lorentz symmetry. This happens because we have not analysed the condition $a_{n+1}=0$ given in Eq. (\ref{2.10}) yet. In order to analyse the condition $a_{n+1}=0$, let us assume that the parameter $\eta$ can be adjusted in such a way that the condition $a_{n+1}=0$ is satisfied. The fact behind this condition is that it yields a relation between the parameter $\eta$ and the quantum numbers $\left\{n,\,l\right\}$ of the system \cite{bb2,bm,eug}. From the mathematical point of view, the relation of the parameter $\eta$ to the quantum numbers $\left\{n,\,l\right\}$ results from the fact that polynomial solutions to Eq. (\ref{2.6}) are achieved for some values of the parameter $\eta$. This means that not all values of the parameter $\eta$ are allowed, but some specific values of $\eta$ that depend on the quantum numbers $\left\{n,\,l\right\}$; thus, we label $\eta=\eta_{n,\,l}$. From the quantum mechanics point of view, this is a quantum effect associated with the influence of the Lorentz symmetry breaking effects on the bound states solutions of this system. In this way, the two conditions established in Eq. (\ref{2.10}) are satisfied and a polynomial expression for the function $H\left(\zeta\right)$ given in Eq. (\ref{2.7}) is obtained. As an example, we consider the ground state $\left(n=1\right)$ and analyse the condition $a_{n+1}=0$. For $n=1$, we have $a_{2}=0$, then, we obtain
\begin{eqnarray}
\eta_{1,\,l}=\frac{\left(2m\left|l\right|-\delta+m\right)\left(2m\left|l\right|+3m-\delta\right)}{2\left(2\left|l\right|+1\right)}
\label{2.12}
\end{eqnarray}

Analyses of parameters that depend on the quantum numbers of the system have been made in recent years. For instance, a relation involving the harmonic oscillator frequency, the Lorentz symmetry-breaking parameters and the total angular momentum quantum number has been obtained in Ref. \cite{bb2}. A relation involving the mass of a relativistic particle, a scalar potential coupling constant and the total angular momentum quantum number has been achieved in Ref. \cite{eug}. In Ref. \cite{bm}, a relation involving a coupling constant of a Coulomb-like potential, the cyclotron frequency and the total angular momentum quantum number in semiconductors threaded by a dislocation density has been obtained.

Next, by considering the simplest case of the function $H\left(\zeta\right)$ which corresponds to a polynomial of first degree ($n=1$), we can write
\begin{eqnarray}
H_{1,\,l}\left(\zeta\right)=1+\frac{1}{\sqrt{\eta_{1,\,l}}}\,\frac{\left(2m\left|l\right|+m-\delta\right)}{\left(2\left|l\right|+1\right)}\,\zeta.
\label{2.14}
\end{eqnarray}
Thereby, from Eq. (\ref{2.14}), thus, the radial wave function given in Eq. (\ref{2.5}) can be written in the form: 
\begin{eqnarray}
R_{1,\,l}\left(\xi\right)=e^{-\frac{\zeta^{2}}{2}}\,e^{\frac{m}{\sqrt{\eta}}\,\zeta}\,\zeta^{\left|l\right|}\,\left(1+\frac{1}{\sqrt{\eta_{1,\,l}}}\,\frac{\left(2m\left|l\right|+m-\delta\right)}{\left(2\left|l\right|+1\right)}\right)\,\zeta.
\label{2.15}
\end{eqnarray}

Hence, the general expression for the energy levels (\ref{2.11}) must be written as:
\begin{eqnarray}
\mathcal{E}_{n,\,l}^{2}=2\,\eta_{n,\,l}\,\left[n+\left|l\right|+1\right].
\label{2.13}
\end{eqnarray}

By observing the confinement of the relativistic scalar particle to a linear scalar potential introduced by a coupling with the mass term, one can expect that the ground state would be defined by the quantum number $n=0$. However, the effects of the Coulomb-like potential induced by Lorentz symmetry breaking effects on the relativistic scalar particle confined to a linear scalar potential yield a change of the energy levels, where the ground state is defined by the quantum number $n=1$. Moreover, a quantum effect characterized by the dependence of parameter $\eta$ of the linear confining potential on the quantum numbers $\left\{n,l\right\}$ of the system arises, and its meaning is that not all values of the parameter $\eta$ are allowed in order to obtain the bound states solutions. Therefore, the conditions established in Eq. (\ref{2.10}) are satisfied and a polynomial solution to the function $H\left(\zeta\right)$ given in Eq. (\ref{2.7}) is achieved in agreement with Refs. \cite{vercin,mhv,eug,b50}.

\section{conclusions}

We have investigated the effects of a Coulomb-type potential induced by a Lorentz symmetry violation background on a relativistic scalar particle. Despite being inspired by the CPT-even sector of the SME, our proposal is not inserted in the scenario of SME, which can be promising because it can open new discussions about new theories that covers energy scales beyond the SME. Since terms that violate the Lorentz symmetry start with tiny values and go for energy scales beyond the Standard Model, we have relaxed the renormalization property of our model. Thereby, we have shown that bound states solutions to the Klein-Gordon equation can be obtained in a particular scenario of the Lorentz symmetry violation defined by a radial electric field, a uniform magnetic along the $z$-direction and tensor background that governs the Lorentz symmetry violation possessing a non-null component given by $\left(\kappa_{DB}\right)_{13}=\kappa=\mathrm{const}$. In this case, we have obtained two allowed energy levels for the system.

Our second discussion has dealt with the effects of a Coulomb-type potential induced by a Lorentz symmetry violation background on the confinement of the relativistic scalar potential to a linear confining potential. In this case, we have introduced the linear scalar potential as a modification of the mass term of the Klein-Gordon equation. We have shown that the effects of the Coulomb-like potential yields the ground state of the system to be defined by the quantum number $n=1$ instead of the quantum number $n=0$, and it also yields the appearance of a quantum effect characterized by the dependence of parameter $\eta$ of the linear confining potential on the quantum numbers $\left\{n,l\right\}$ of the system. The meaning of this dependence is that not all values of the parameter $\eta$ are allowed in order to obtain the bound states solutions.

\acknowledgments

The authors would like to thank CNPq (Conselho Nacional de Desenvolvimento Cient\'ifico e Tecnol\'ogico - Brazil) for financial support.

\end{document}